# DNA Hash Pooling and its Applications[*]

Dennis Shasha[♦]
Courant Institute of Mathematical Sciences
New York University
USA
shasha@cs.nyu.edu

Martyn Amos
Department of Computing and Mathematics
Manchester Metropolitan University
UK
M.Amos@mmu.ac.uk

**Abstract**

In this paper we describe a new technique for the comparison of populations of DNA strands. Comparison is vital to the study of ecological systems, at both the micro and macro scales. Existing methods make use of DNA sequencing and cloning, which can prove costly and time consuming, even with current sequencing techniques. Our overall objective is to address questions such as: (i) (*Genome detection*) Is a known genome sequence present, at least in part, in an environmental sample? (ii) (*Sequence query*) Is a specific fragment sequence present in a sample? (iii) (*Similarity discovery*) How similar in terms of sequence content are two unsequenced samples? We propose a method involving multiple filtering criteria that result in   pools   of DNA of high or very high purity. Because our method is similar in spirit to hashing in computer science, we call it *DNA hash pooling*. To illustrate this method, we describe protocols using pairs of restriction enzymes. The *in silico* empirical results we present reflect a sensitivity to experimental error. Our method will normally be performed as a filtering step prior to sequencing in order to reduce the amount of sequencing required (generally by a factor of 10 or more). Even as sequencing becomes cheaper, an order of magnitude remains important.

# INTRODUCTION

Biologists often examine large and diverse populations of organisms (for example, molecules, microbes or plants). This is particularly the case in fields such as microbial ecology, which studies the interactions between living microorganisms (such as algae, or bacteria) and their environment. One of the most significant and challenging problems in these areas of biology is to compare the species in different samples. This task is often made even more difficult by the fact that many   wild   organisms resist laboratory cultivation (and, thus, have unknown phenotypes and their genomes are unknown), or may be present in a population in relatively low numbers.

The study of *metagenomics* has emerged in recent years (Zhou, 2003; Handelsman, 2004; Tringe *et al.*, 2005; Huson *et al.*, 2007) to perform what has been described as "environmental forensics", including the quantification of relative abundances of known species, and the estimation of the number of "unknown" species in a given environment (Huson *et al.*, 2007). The potential impact of this new field is huge, with applications ranging from medicine to agriculture and biotechnology. Furthermore, the insights gained will be of significant assistance in advancing our understanding of biodiversity in both new and familiar environments, such as frozen Antarctic lakes and the human gut (Handelsman, 2004).

The rest of this paper is organized as follows: We provide a brief overview of metagenomics, describing current techniques and practices. We discuss several shortcomings of existing methods, motivating the work presented in the following section. Here, we present our new method for population analysis, along with the results of *in silico* experiments. We conclude with a discussion of the implications of this work, and suggest possible future lines of enquiry.

## AN OVERVIEW OF METAGENOMICS

The term *metagenomics* was first used in print by Handelsman *et al.* (Handelsman *et al.*, 1998) to describe the study of genetic information recovered from environmental samples. In line with the other "omics" (e.g., proteomics, metabolomics), the emphasis lies on *sets* of products and/or genes, the suffixes "-ome" and "-omics" being interpreted to imply *totality* or *collectivity* (Lederberg and McCray, 2001).

According to Schloss and Handelsman, "metagenomics is the culture-independent analysis of a mixture of microbial genomes (termed the metagenome) using an approach based either on expression or sequencing" (Schloss and Handelsman, 2005). Only a tiny fraction (estimated at below 1%) of micro-organisms found in nature are amenable to isolation and culture for further study (Ammann *et al.*, 1995). This means that we need alternative ("culture-independent") strategies for studying the vast majority of microbes found in the wild, a significant number of which will be unknown to science (Ward *et al.*, 1990).

The possible implications of metagenomics are many and wide-ranging. As Handelsman argues, "Microbiology has long relied on diverse methods for analysis, and metagenomics can provide the tools to balance the abundance of knowledge attained from culturing with an understanding of the uncultured majority of microbial life. Myriad environments on Earth have not been studied with culture-independent methods ... and they invite further analysis. Metagenomics may further our understanding of many of the exotic and familiar habitats that are attracting the attention of microbial ecologists, including deep sea thermal vents, acidic hot springs, permafrost, temperate, desert, and cold soils, Antarctic frozen lakes, eukaryotic host organs, the human mouth and gut, termite and caterpillar guts, plant rhizospheres and phyllospheres, and fungi in lichen symbioses. With improved methods for analysis ...and attraction of diverse scientists to identify new problems and solve old ones, metagenomics will expand and continue to enrich our understanding of microorganisms" (Handelsman, 2004).

## Methods for Metagenomics

Early efforts to characterise environmental samples used *ribosomal RNA* (rRNA). This class of RNA plays a major role in protein synthesis, and rRNA makes up at least 80% of the RNA molecules in a typical eukaryotic cell. The gene encoding one particular molecule - 16S rRNA - is *non-coding* - rather than "becoming" (or being *translated* into) a protein, the RNA molecule resulting from its transcription "assists" in protein synthesis. The gene is *highly conserved*, meaning that it has been preserved throughout the process of speciation through evolution (the implication being that it performs a fundamental role in basic biochemical processes that are essential for life). This conservation across species gives 16S rRNA a degree of *universality*, since it is present in a diverse range of organisms.

However, despite its highly conserved nature, 16S rRNA also has the property of *hypervariability*, which also makes it useful to biologists. The average 16S rRNA molecule is 1,500 nucleotides long, which gives ample space for variation between (groups of) organisms. The *evolutionary distance* between any two organisms can therefore be calculated by aligning their 16S rRNA sequences (for example) and calculating the number of differences (Pace, 1997). We can use these *known* gene sequences to identify microbes in an environmental sample. Any *unknown* variants of the gene must therefore correspond to unknown organisms.

Early work on using 16S rRNA to study microbial ecology was performed by Pace and colleagues (Lane *et al.*, 1985; Olsen *et al.*, 1986) (see also (Head *et al.*, 1998)), and further advances came with the development of the polymerase chain reaction (PCR) (Mullis and Faloona, 1987). The universality of 16S rRNA means that if we wish to amplify (and thus clone) DNA from a population of microbes, we may use a pair of sequences encoded within it as primers for PCR, without advance knowledge of which specific organisms are present in the sample (Medlin *et al.*, 1988; Weisburg *et al.*, 1991; A.L. *et al.*, 1992).

Although 16S rRNA sequencing is a powerful and well-used technology, doubts have been raised over its resolution and absolute applicability (Janda and Abbott, 2007). By amplifying a sample, we may drown out relatively rare sequences, may not differentiate between subspecies that have common ribosomal RNAs but differ elsewhere (perhaps by incorporating foreign DNA easily, as is the case for *Bacillus subtilis*), and will miss viruses (which do not contain the gene).

Until recently, metagenomic analysis involved the extraction of DNA from an environmental sample, cloning of the DNA into a suitable vector, insertion of the vector into a host bacterium and then screening the resulting transformed bacteria (Handelsman, 2004). Screening may occur on the basis of *gene expression* using microarrays (Zhou, 2003), using some other trait, such as antibiotic production (Schloss and Handelsman, 2003), or simply via sequencing.

A combination of recent advances (which are largely computational) have changed the landscape substantially. Shotgun sequencing (Fleischmann *et al.*, 1995) has been applied with great success to metagenomics. This method of sequencing randomly shears long DNA strands into shorter fragments, which are then individually sequenced and then reassembled *in silico* into a consensus sequence.

The assembly of genomes from complex communities historically demands enormous sequencing expenditure for the assembly of even the most predominant members (Tringe*et al.*, 2005). Because of this difficulty, borne out by initial studies by Tringe *et al.* (Tringe *et al.*, 2005), the authors decided to employ an alternative, gene-centric approach that does not attempt to attribute genes obtained to any particular genome. They obtained their initial dataset by taking four sets of samples, one from agricultural soil, and three from whale carcasses. Samples were then partitioned into bacteria, archea or eukaryotes using PCR-amplified rRNA libraries. Genomic small-insert libraries were then shotgun sequenced from each sample (100 million base pairs from the soil and 25 million base pairs from each whale sample). These sequences, derived from different population members, were termed Environmental Gene Tags (EGTs), since they may encode regions of functional genes that are necessary for survival in a particular environment. Different environment types will exhibit unique EGT fingerprints, containing genes derived from many different genomes. The study showed that two whale carcasses, located 8000km apart, nontheless had very similar EGT patterns. Thus, one may determine the type of environment from this fingerprinting technique.

Other studies of note that have applied shotgun sequencing to metagenomics include an examination of the ecology of the Sargasso Sea (Venter *et al.*, 2004) and drainage in an acid mine (Tyson *et al.*, 2004). For a brief introduction to the field and a review of recent work, the reader is directed to (Eisen, 2007).

In a recent study, Huson *et al.* present an approach (which they call MEGAN, for MEtaGenome ANalyzer) (Huson *et al.*, 2007) to the problem of genomic assembly in which the authors compare sequenced data to existing databases. Specifically, the set of DNA sequences obtained by random shotgun sequencing from the environmental sample is run against known sequences using BLAST. The resulting meta-data is then provided as input to the MEGAN package, which estimates and explores the taxonomical content of the data set. This may be a good technique to obtain the most abundant species in a sample, but will have difficulty locating rare sequences of interest. One of the themes of our proposed approach is to hunt systematically for signs of a genome of interest.

Even for a relatively simple community study on the drainage region of an acid mine, roughly 15 million bases were sequenced in order to obtain the required metagenomic data (Tringe *et al.*, 2005), at a cost of approximately $150,000. A soil study, requiring at least 50 million bases, might then cost half a million US dollars. Recent developments such as *pyrosequencing* (Ronaghi, 2001; Elahi and Ronaghi, 2004; Ahmadian *et al.*, 2006) and other high throughput sequencing efforts have reduced the cost of sequencing substantially, facilitating large-scale metagenomics (Mardis, 2008), such as the recent study of catastrophic collapses of honey bee colonies (Cox-Foster *et al.*, 2007). This emerging technology permits groups to consider for the first time, hugely ambitious projects, such as sequencing the human *biome* (the entire ecosystem of the human body) (Blow, 2008).

These are still substantial efforts that attempt to understand an entire ecosystem. However, such efforts may not be necessary in order to answer potentially simpler questions such as: What is in common between these two samples? In the next section, we consider the notion of pre-processing a sample in order to reduce the complexity of any analysis.

## SAMPLE PRE-PROCESSING

Fortunately, sequencing is not always necessary as a first step. Molecular techniques that work at the whole sequence level may be used to reduce the initial complexity of a sample population. One tool commonly employed is GC fractionation (Holben and Harris, 1995), which works along the principles of a molecular sieve, sorting strands according to their relative GC content (guanine and cytosine being heavier than their counterparts adenine and thymine). This may be effective when trying to partition a sample into *eukaryotic* and bacterial sets, since eukaryotic DNA tends to have a much lower GC content (e.g, we selected two complete bacterial genome sequences, A (*Escherichia coli* K12) and B (*Shigella boydii* Sb227) for early studies; each of these had a GC content of roughly 51%, while the human genome is made up of around 45% GC and that of the mouse roughly 44%). However, such a relatively crude tool rapidly proves ineffective when dealing with shorter sequences, where we may only possess genomic fragments within our sample. For bacterial sequence A, when taking 200 random consecutive sequences of length 50,000, we obtained a GC content ranging from 46.7% to 53.3% with the 90% confidence interval ranging from 47.3% to 52.8%.

Preliminary work on estimating the complexity of a heterogenous population of DNA strands (without using sequencing) is reported in (Faulhammer *et al.*, 1999).

This paper, motivated in part by the authors' earlier work on DNA-based computing (Adleman, 1994), reports initial experimental investigations into the use of basic laboratory methods (combined with probability theory) to estimate the complexity of a tube of strands. Faulhammer *et al.* digested their initial tube with a set of restriction enzymes with recognition sites differing in sequence and of length $4 \leq k \leq 8$. The contents of the tube were then visualised in a gel, and the *number* of distinct bands observed used to obtain an estimate of the number of different strands.

This basic approach (the use of restriction enzymes to digest a population sample, followed by analysis of the fragment size) also underpins an early variant of the well-known technique of *DNA fingerprinting* (Jeffreys *et al.*, 1985). *Restriction fragment length polymorphisms* (RFLPs) (Dowling *et al.*, 1990) provide a technique by which organisms may be differentiated by comparing the patterns obtained by digesting a certain portion of their DNA. If two organisms differ in the distance between restriction sites, the length of the fragments produced will differ (i.e., be polymorphic) when the DNA is digested. However, this method is generally only useful when the population sample is relatively homogeneous (e.g., one wishes to distinguish between members of the same species).

A related technique, involving amplification of 16S rRNA followed by restriction enzyme digestion, has been used to detect pathogens in spinal fluid (Lu *et al.*, 2000) and characterize the diversity of model communities (Liu *et al.*, 1997). Our proposed technique uses some of the same basic laboratory methods, but it differs from all of the previous approaches in several important ways:

1. We avoid reliance on 16S rRNA.
2. Depending on the exact problem, our process can be simulated *in silico* of performed *in vitro*.
3. We use multiple levels of restriction enzyme digestion to achieve reasonable purity.

In the rest of this paper we present and evaluate a simple and powerful technique called *DNA hash pooling*. We conclude with a discussion of plans for future theoretical and experimental work.

# DNA HASH POOLING

In computer science, *hashing* (Knuth, 1998) maps a relatively small set from a large domain (e.g., 10,000 integers ranging in value from 0 to one billion) to a small domain (e.g., the set of integers from 1 to 5000) through a mathematical *hash function*. Applications of hashing include cryptography, error correction, authentication and identification. A typical hash function is *modulus* (i.e., remainder). For example, 7 mod 5 = 2 because 2 is the remainder after dividing 7 by 5. For the same reason, 28 mod 5 = 3. A hash data structure based on "mod 5" will map 28 to bucket (or *pool*) 3, 7 to pool 2, 12 to pool 2, 59 to pool 4, and so on. There are many variants of hashing, some of which entail hashing each pool resulting from the first hash function in order to get "purer" pools, and then using the combined hash results to generate an item "label". For example, using a second hash function, based on "mod 7", 28 would map to 0 and 53 to 5. Thus the full "label" of 28 would be (3, 0) because 28 mod 5 = 3 and 28 mod 7 = 0. By contrast, the label of 53 would be (3, 4) because 53 mod 5 = 3 and 53 mod 7 is 4. Associated with each unique label is a pool having a relatively small number of distinct values.

*DNA hash pooling* or *hash pooling* for short is the analogous operation on DNA. The "hash functions" in this scenario correspond to biological operations that give rise to distinctive and quantifiable "fingerprints" (e.g., measurement of GC content followed by digestion by a set of restriction enzymes). The label components correspond to the values obtained by application of the hash functions (e.g. GC content and fragment length).

*In silico*, our method involves simulating these operations on known sequences (typically though not necessarily of entire genomes) and characterizing different portions of those sequences from the result(s). *In vitro*, our method involves performing the bench-based operations and sequencing only those pools that are likely to be of interest.

For concreteness, this paper focusses on hash pooling based solely on *restriction enzymes*. This focus is for illustration purposes only. Any experimental technique that separates portions of DNA deterministically into buckets would suffice, including partial sequencing and hybridization.

The basic operations for restriction enzymes are

1. Apply a six base-pair (base pair) restriction enzyme to a sequence, yielding a set of fragments.
2. Partition those fragments based on length (perhaps approximately), using a technique such as gel electrophoresis.
3. Apply a four base pair restriction enzyme to a selected subset of partitions and separate on length again.
4. Sequence selected lengths.

Each resulting pool is therefore associated with a label consisting of two lengths, the first based on a six base pair restriction enzyme and the second based on a four base pair restriction enzyme (Figure 1 ).

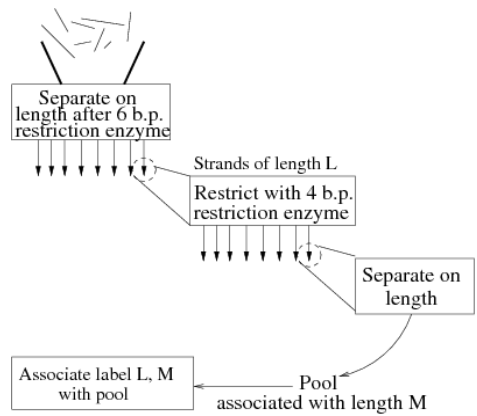

Figure 1: Two-stage hash pooling

The procedure may be described for K stages in pseudo-code as follows:

```
hash(stage j, sample s, label L, K)
  r: = restriction enzyme for stage j
  frags := apply r to s
  mysamps := partition frags on length
  for each t in mysamps
       tlabel := L concat (r, length(t))
       if (j < K)
            hash(j+1, t, tlabel, K)
       else (t, tlabel) is a member of the final pool
  end for

Pseudo-code for K stage hash pooling based on restriction enzymes. The
initial call on an initial  sequence orig would be hash(1, orig, null,
K), where null is the empty label.
```

For example, consider the genomic sequence of bacterium A (*E. coli* K12). If we cut A using the enzyme *Sma*I (recognizing CCCGGG), take the pool corresponding to length 264, cut that pool with *Rsa*I (GTAC) and take the pool of length 31, we get a pool having label (264, 32). It happens to have a single member with the sequence CTATCCGCTCAATGAGTCGGTCGCCATTGCCC. By contrast, the pool with label (770, 207) has three different sequences. For some applications, we will want pools having singletons (i.e., a set with only a single element) in order to obtain a pure sequence without the need for cloning.

One may object that separating fragments by length entails a certain inaccuracy imposed by the laboratory technique; a reasonable estimate of this error may be plus or minus 10 base pairs, if we assume the use of specific sorting techniques (Heiger *et al.*, 1990). In this case, in order to obtain a pure sample, we may be interested in finding a pool whose label has no 10 base pair-neighbors. The labels $L$ and are 10 base pair-neighbors if (i) the first component of $L$ and the first component of are different but differ by 10 or less; or (ii) the first component of $L$ and are the same but the second components differ by 10 or less. For *E. coli* K12, the labels (188, 59) and (188, 106), for example, have no 10 base pair-neighbors. Experimentalists have told us different stories about accuracies. Some think that the 10 base pair criterion is reasonable provided one works carefully and slowly. Others think it is not. We have rerun some of the experiments below assuming that we could measure lengths up to an accuracy of 50 base pairs. The results do not substantially change.

## EXPERIMENTS

Having presented our formal framework, we can now present several applications and our *in silico* empirical results.

## Genome Detection

The first question we ask is the following: given a tube, $T$, of unknown DNA (perhaps from an environmental sample) and a genome whose sequence is known, are reasonably sized portions of that genome present in $T$, even if in small concentrations? (Figure 2 )

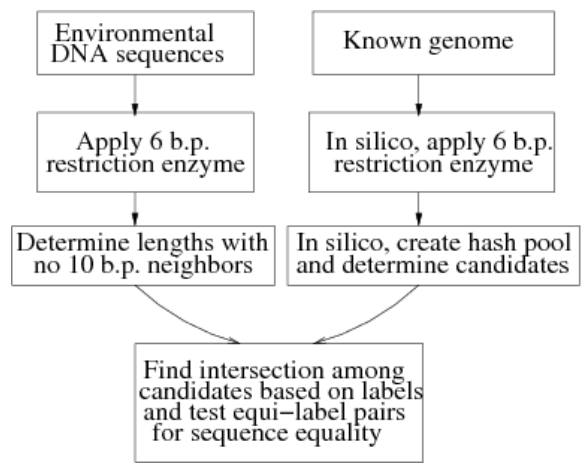

Figure 2: Genome sequence detection

A reasonably sized portion is a sequence of length at least 200,000 base pairs (or roughly 5% of the length of a bacterial genome.) This might be used for the detection of bacterial pathogens in food, for example. In what follows, we used bacterium A, *E. coli* K12, which is often used as an indicator organism in the detection of faecal contamination.

## Genome Detection A

Our *in silico* experiment involved the following steps:

1. Compute the *candidate set* of A, consisting of possibly non-singleton pools having no 10 base pair neighbors. There were 3,567 candidates. This gives us a comparison library of pools. It is important to note that this step is purely computational and can be computed just once for any combination of known genome and restriction enzyme set.
2. We simulate the unknown sample $T$ by taking a 200,000 consecutive base pair subsequence of A (with the start position taken uniformly at random) and combining it with a sequence of length four times that of A (generated pseudo-randomly to have the same GC content as A).
3. We then compute the resulting candidate set of hash pools, based on no 10 base pair neighbors.

Having tried this 20 times, we found, on average, 2,000 pools in the second candidate set. On average the size of the the two sets had common labels only 71 times. When labels were equal, 99.8% of the time there was a match of the sequence *and* the sequence came from that 200,000 consecutive base pair subsequence, giving a precision of 99.8% (only 3 times out of 1436 were the labels the same but the sequences different over the 20 attempts). Of the 200,000 base pairs in the common sequence, more than 24% are found (standard deviation 9%). This required sequencing under 50,000 base pairs on the average with a standard deviation of under 20,000. Since the test sequence is over 18 million base pairs long, this is more than a factor 100 reduction in necessary sequencing.

When applying this in a laboratory setting, there is the significant question of whether this operation may require many separate DNA extractions and applications of a restriction enzyme. Fortunately, the answer is no. For each of the 20 tests, first the six base pair restriction enzyme was used. This gave a collection of fragment lengths. On average only 5.8 of those lengths had no 10 base pair neighbors and had lengths similar to the lengths of the candidates from A (typical lengths were between 7,000 base pairs and 39,000 base pairs). So on average only 5.8 fragment lengths required extraction. Of those, 4.7 (on average) yielded matching sequences. So, if this were done *in vitro*, approximately 70 common strands would be found using one application of *Sma*I and under six applications of *Rsa*I. Virtually all (99.8%) tested strands would be shown to be equal.

## Genome Detection B

A variation of experiment Genome Dection A is to embed the 200,000 consecutive base pair subsequence of A into a related bacterium B (*Shigella boydii* Sb227). In that case, when labels were equal, 33% of matching labels (1921 out of 5877 matching labels) led to matching sequences among the 200,000 consecutive base pair subsequence. This method gives reasonable coverage: an average of 62,000 base pairs among the 200,000 base pairs are covered (standard deviation of 11,000). Total sequencing costs were again modest: 146,000 on the average with a standard deviation of 10,500.

As a further variation, consider the case where the 200,000 consecutive base pair subsequence of A to be placed in B as above, but length measurement is accurate to only 50. This means that we would not consider labels unless they had no 50 base pair neighbors. This gave 100% precision, but found only 136 matching labels. The coverage was 4.3% (8,600 base pairs found) with a standard deviation of 3.1. Only 8,600 base pairs were sequenced as well with the same standard deviation. This shows that less accurate read lengths can reduce coverage but preserve (in fact increase) precision.

This experiment shows that *in silico* hash pooling on a known genome can identify pools to look for in a sample, such that those pools have a strong likelihood of containing a subsequence of the known genome. Thus, we can see this method as an improvement over random sampling, and can be used even if the bacterium of interest is relatively rare in the sample.

## Sequence Query

Here we address a closely related question to our first experiment: given a query sequence, is that sequence present, at least *in part*, in the tube? This might be used to look for the presence of a pathogen, for example. In fact, an experiment similar to that used to address the first question allows us to measure the effectiveness of our approach.

Suppose the sample under scrutiny contains A plus a lot of other assorted DNA (e.g. the full genome of A amongst a pseudo-random sequence four times the length of the A sequence and having the same GC content). Then, 20 times, we take a random query subsequence of length 200,000 from A and see if we can find matching parts in the sample tube.

The sample tube (A sequence plus a *random* sequence four times A in length with no 10 base pair neighbors) has 3,516 candidate pools. The average 200,000 base pair subsequence of A has about 200 candidates. In our 20 experiments, whenever two labels are equal, the corresponding sequences matched 100% of the time (precision of sequence matching given label match of 100%). This is not guaranteed to hold always of course, but again shows that even without sequencing one can be quite sure that sequences will match if labels match.

As in the first experiment, the six base pair restriction enzyme would cut the fragments into certain lengths, but, on the average, only 2.3 of those lengths (ranging from 10,000 base pairs to 30,000 base pairs) would have the properties that (i) they had no 10 base pair neighbors and (ii) they matched the candidates from the 200,000 base pair query sequence. Thus, on the average, under three extractions need to be taken and then digested by the four base pair restriction enzyme.

Consider now the negative case when the query sequence was nowhere present in the sample, On the average, after cutting with the six base pair restriction enzyme, on the average, under one of those lengths had the properties that (i) they had no 10 base pair neighbors and (ii) they matched the candidates from the 200,000 base pair query sequence. When extracted and digested by the four base pair restriction enzyme, there were no matching labels (other than a single label whose final fragment length was only 4). So this technique does not throw up false positives.

## Similarity Discovery in the Field

Here we consider the problem: given two tubes of DNA, do they contain strands that are the same or very similar? This might be useful when comparing samples of unsequenced genomes, but rRNA genes are nor enough. In this case, we cannot compute candidate pools that have no 10 (or 50) base pair neighbors using known genomes. Instead, we have to measure them. Sometimes we may not know whether the sample contains known sequences. If it does, we can use the techniques in the Subsequence Detection subsection above to find out which known genomes each sample contains and then see which are the same.

Let us assume, however, that the sample contains no known genomes (or that we want to detect commonalities *besides* those among known genomes). Our strategy will be to choose the most *likely* pairs to study by focussing on unusual labels (Figure 3). We therefore performed the following *in silico* experiment 20 times:

1. Take a 200,000 base pair sequence, *target*, with the same GC content as A, plus a random sequence four times the size of A (4×4.7*Mb*=≈20Mb), with the same average GC content as A.
2. For the second sample, we use the same 200,000 base pair sequence *target* plus *another* random sequence four times the size of A, with the same average GC content. Thus the target in each sample is 200,000 base pairs long, just 1% of the roughly roughly 20 million for the entire sequence present in each tube.

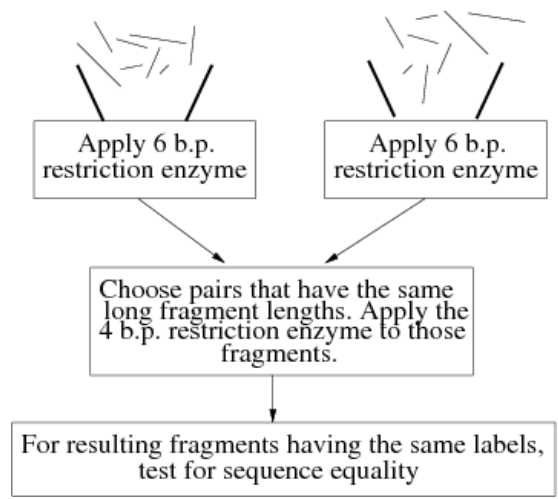

Figure 3: Sample comparison

In which pools should we look for common strands? That is, is it better to look at pools where the six base pair restriction enzyme has cut strands of length approximately 4,000 (the expected value) or much longer? The *in silico* answer is obvious in retrospect: to find common strands, the best pools to look at are ones corresponding to *long* lengths when cut by the first restriction enzyme. Thus the procedure is this:

1. Cut each sample with the six base pair restriction enzyme, then find all lengths that are the same (within an accuracy of 10 base pairs).
2. On the upper quartile of those lengths (approximately 235 of them), apply the four base pair restriction enzyme.

Of those fragments that have the same lengths (within 10 base pairs) for both the first and second restriction enzymes, between 4% and 7% are the same sequence over the 20 experiments that we tried. Other quartiles are about a factor of 10 less good.

An even better approach is to look at dectiles. Considering only the upper 1/10 of the lengths from the 6-base pair restriction site (from 10,210 to 24,550) gives a hit ratio of 13% on the average (standard deviation of 2%) and only 94 lengths (standard deviation under 1) from the six base pair restriction enzyme require an application of the second restriction enzyme. The upper dectile covers an average of 18% of the 200,000 base pair common string (standard deviation 6%). The upper dectile requires sequencing only about 330,000 bases on the average (standard deviation 18,000). Since sequence A is 4.5 million base pairs by itself, this represents more than a factor of 10 reduction in sequencing.

If we have already identified known genomes that the two tubes share in common, then we should avoid labels that correspond to those. This allows us to avoid resequencing known commonalities.

**Field Similarity B**

As above, we take a 200,000 base pair sequence, *target*, with the same GC content as A, plus a random sequence four times the size of A ($4 \times 4.7Mb = \approx 20Mb$), with the same average GC content as A. For the second sample, we use the same 200,000 base pair sequence *target*, but *embed it into bacterium B*. The same method works. In this case, the upper dectile yields a percentage of hits (successful sequence matches/label matches) of 62% on the average with a standard deviation of 16%. That is, the high dectiles gives even higher precision when embedding the target in sequence B ((*Shigella boydii*) than in a random sequence. This comes at a cost of coverage. The upper dectile covers an average of only 12% of the 200,000 base pair common string (standard deviation 4%). The upper dectile requires sequencing only 70,000 bases on the average (standard deviation 14,000). This is a factor of 100 reduction of sequencing compared with sequencing the two genomes.

When the samples are large (contain many different genomes) and the results of the cuts by the larger restriction enzymes show no separation, the best labels are the ones that have few neighbors (the long cuts from the 6 base pair restriction enzymes). To determine whether a pool from one sample might match a pool having the same good" label as one from the other label, one could hybridize the two samples. If they match, then it is worthwhile to sample them.

## CONCLUSIONS

DNA hash pooling is a method to simplify many problems in metagenomics. It gives the experimenter the ability to query for known sequences and genomes in a sample or to find common sequences from unknown genomes in two or more samples *even if the identified sequences are rare*. The version of the technique described in this paper involves a small number of steps of the form: extract DNA of a certain length, apply a restriction enzyme to it, and measure the lengths of the results. In most cases, a few cuts with restriction enzymes can reduce sequencing to the best candidates.

The main future work we anticipate is to validate the technique and then extend the method as new application scenarios present themselves.

## ACKNOWLEDGEMENTS

Shasha s work has been partly supported by the U.S. National Science Foundation under grants IIS-0414763, DBI-0445666, N2010 IOB-0519985, N2010 DBI-0519984, DBI-0421604, and MCB-0209754. This support is greatly appreciated. We are also grateful to Laura Landweber, Rob Knight and David A. Hodgson for helpful advice and discussions.